\documentclass[aps,prl,twocolumn,secnumarabic]{revtex4-1}
    \usepackage{graphicx}
    \usepackage{amsmath}

    \newcommand{\bs}{\boldsymbol}

\begin{document}
        
        \title{Universal Dissipationless Dynamics in Gaussian Continuous-variable Open Systems}
        \author{Han-Jie Zhu$^{1}$}
        \author{Guo-Feng Zhang$^{1}$}
        \email{gf1978zhang@buaa.edu.cn}
        \author{Lin Zhuang$^{2}$}
        \author{Wu-Ming Liu$^{3}$}
        \affiliation{$^{1}$School of Physics and Nuclear Energy Engineering, Beihang University, Beijing, China}
        \affiliation{$^{2}$State Key Laboratory of Optoelectronic Materials and Technologies, School of Physics, Sun Yat-Sen University, Guangzhou, China}
        \affiliation{$^{3}$Beijing National Laboratory for Condensed Matter Physics, Institute of Physics, Chinese Academy of Sciences, Beijing, China}
        \begin{abstract}
            We investigate the universal dissipationless dynamics of Gaussian continuous-variable systems 
            in the presence of a band-gapped bosonic environment. 
            Our results show that environmental band gaps can induce localized modes, 
            which give rise to the dissipationless dynamics 
            where the system behaves as free oscillators instead of 
            experiencing a full decay in the long time limit. 
            We present a complete characterization of localized modes, 
            and show the existence of the critical system-environment coupling. 
            Beyond the critical values, localized modes can be produced 
            and the system dynamics become dissipationless. 
            This novel dynamics can be utilized to overcome the environmental noises and 
            protect the quantum resources 
            in the continuous-variable quantum information.
        \end{abstract}
        \maketitle

        The dissipation and decoherence processes induced by surroundings are the central topic 
        of study in the theory of open quantum systems. 
        These unavoidable processes almost always lead to 
        the irreversible loss of quantum coherence and quantum correlations 
        which are the crucial resources for quantum technologies 
        \cite{Dynamics_of_non_Markovian_open_quantum_systems,Natural_and_artificial_atoms_for_quantum_computation,Quantum_sensing,Microwave_photonics_with_superconducting_quantum_circuits,Atomic_physics_and_quantum_optics_using_superconducting_circuits,Recent_developments_in_trapping_and_manipulation_of_atoms_with_adiabatic_potentials}. 
        Yet, quantum resources may be protected by utilizing quantum states 
        that are less sensitive to environmental perturbations. 
        This constitutes the basic idea of passive protection of quantum resources, 
        which has been an important subject and had many applications in quantum information \cite{RevModPhys.88.041001.4}.
        A prominent example is that of the decoherence-free subspaces, 
        where the state evolution inside these subspaces is 
        completely unitary \cite{zanardi1997noiseless5,kielpinski2001a6,Correlation_induced_suppression_of_decoherence_in_capacitively_coupled_Cooper_pair_boxes,Protected_quantum_state_transfer,Single_shot_realization_of_nonadiabatic_holonomic_quantum_gates,Decoherence_Free_Interaction_between_Giant_Atoms_in_Waveguide_Quantum_Electrodynamics}. 
        Another novel example is the bound state 
        where the discrete eigenstate is formed inside the environmental band gap. 
        Such state is stable under environmental noise, 
        and gives rise to dissipationless dynamics when the non-Markovian effect is fully taken into account. 
        For finite dimensional systems such as spin systems, 
        this phenomenon can occur for systems embedded in 
        photonic band-gap materials, 
        and is known as atom-photon bound states 
        \cite{john1990quantum7,john1994spontaneous8,liu2013anomalous10,PhysRevA.91.062112.11,shen2016quantum12,Scattering_in_the_Ultrastrong_Regime,Nonequilibrium_many_body_steady_states_via_Keldysh_formalism,Threshold_for_formation_of_atom_photon_bound_states,Bound_States_in_Boson_Impurity_Models,Bound_state_in_the_continuum,Suppressed_dissipation_of_a_quantum_emitter,Atom_field_dressed_states,Effect_of_counter_rotating_terms}. 
        This feature can lead to 
        many practical phenomena such as decoherence suppression \cite{tong2010decoherence13}, 
        quantum entanglement and correlation preservation \cite{tong2010mechanism14,lazarou2012entanglement15,Preservation_of_quantum_correlation,Trapping_of_coherence_and_entanglement,Bound_states_and_entanglement_generation,Robust_fermionic_mode_entanglement}, 
        quantum speedup \cite{Mechanism_for_quantum_speedup}, 
        and metrology precision enhancement \cite{1367-2630-19-11-113019.17}. 

        Despite significant progress on the subject, 
        the analysis has almost exclusively been focused on few-body systems. 
        The analogous behavior for many-body quantum systems is of general interest and highly desirable, 
        since ultimately many schemes in practice require such system. 
        However, it is still a challenging and almost unexplored topic to 
        understand such behavior in many-body systems.
        For another category of quantum systems, i.e., continuous-variable (CV) systems, 
        another question arises: can bound states still be formed in the band-gapped environment and if so, 
        do these bound states lead to dissipationless dynamics?
        {\bf
        These questions are especially interesting for Gaussian systems, 
        which constitute a large class of CV systems, 
        and play a central role as well as serve as primary tools in CV quantum information. 
        It would be novel and of practical relevance if bound states can be formed in Gaussian systems, 
        which allows us to protect quantum resources in many CV protocols.
        }
        Despite some previous papers have addressed related simplified problems 
        \cite{yang2014canonical18,an2007non-markovian20,wu2010non-markovian21,PhysRevA.96.023831.22,Coherence_of_mechanical_oscillators,Non_equilibrium_quantum_phase_transition,General_Non_Markovian_Dynamics_of_Open_Quantum_Systems}, 
        the models discussed in these articles are small systems, 
        and can be regarded as finite dimensional systems 
        since oscillators behave as few-level atoms in their invariant subspaces. 
        Thus these models may not provide correct bound-state properties in 
        Gaussian systems, even for the single oscillator case. 
        Therefore, it is essential to establish a bound-state theory 
        for general Gaussian systems. 

        In this Letter, we explore the Gaussian dissipationless dynamics 
        in general CV systems with band-gapped bosonic environments and show that, 
        different from finite dimensional systems, 
        bound states in CV systems are characterized by localized modes 
        with frequencies embedded in environmental spectral gaps. 
        We obtain existence conditions of localized modes by analyzing non-Markovian dynamics, 
        and our results confirm that these modes will give rise to dissipationless dynamics 
        where the system behaves as free oscillators in the long time limit. 
        As an important case, we analyze localized-mode properties 
        in the weak system-environment coupling limit 
        which can be satisfied by most experimental settings. 
        We further illustrate our results in an experimentally achievable system. 
        
\begin{figure}
            \centering
            \includegraphics[scale=0.160]{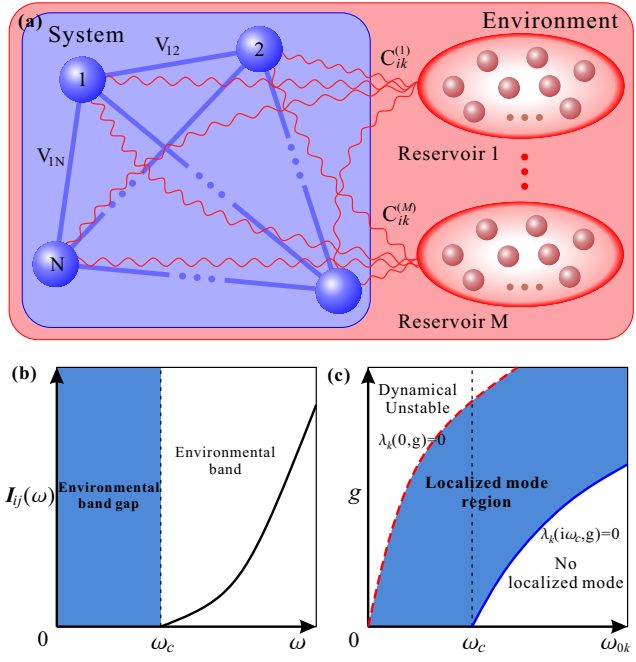}
            \caption{
                (a) Illustration of the general Gaussian network of open oscillators. 
                Here $N$ interacting oscillators {\bf(blue spheres)} are bilinearly coupled with bosonic reservoirs, 
                while each reservoir consists of different noninteracting bosonic modes {\bf(red spheres)}. 
                (b) Band-gapped structure of the environment spectrum. 
                (c) Localized modes status. 
                The $k$th localized mode presents above 
                the critical coupling line (blue solid line). 
                Meanwhile, the system becomes dynamical unstable in the regime 
                which fails to meet the stability condition (above the red dashed line).
            }
            \label{Fig1}
        \end{figure}

        We consider the system consists of $N$ 
        interacting oscillators bilinearly coupled to a general bosonic environment, 
        as shown in Fig.$\,$\ref{Fig1}(a). 
        The total Hamiltonian can be written as $\hat{H} \! = \! \hat{H}_S+\hat{H}_E+\hat{H}_{int}$, 
        where $\hat{H}_S \! = \! \bs{P}^T \bs{P}/2 \! + \! \bs{X}^T \bs{V}\bs{X}/2$ 
        is the system Hamiltonian which describes $N$ oscillators. 
        The $N \times N$ matrix $\bs{V}$ defines the interaction between these oscillators, 
        while column vectors $\bs{X} \! = \! (\hat{x}_1,\hat{x}_2,\ldots,\hat{x}_N)^T$ and 
        $\bs{P} \! = \! (\hat{p}_1,\hat{p}_2,\ldots,\hat{p}_N )^T$ 
        store coordinates and momenta of oscillators. 
        The environment contains $M$ bosonic reservoirs and is described by the Hamiltonian 
        $\hat{H}_E \! = \! \sum_\alpha \hat{H}^{(\alpha)}$ 
        with $\hat{H}^{(\alpha)} \! = \! \sum_k ( \hat{p}_k^{(\alpha)2} + \omega_k^2 \hat{q}_k^{(\alpha)2} )/2$. 
        The interaction Hamiltonian is 
        $\hat{H}_{int} \! = \! \sum_{\alpha ik} C_{ik}^{(\alpha)} \hat{x}_i \hat{q}_k^{(\alpha)} \! = \! 
        \bs{X}^T \sum_{\alpha} \bs{C}^{(\alpha)} \bs{Q}^{(\alpha)}$ 
        where each oscillator can simultaneously interact with many reservoirs, 
        and $C_{ik}^{(\alpha)}$ is the coupling between the $i$th oscillator 
        and mode $k$ in the reservoir $\alpha$.
        This interaction is relevant in many realistic scenarios 
        where the oscillator is principally coupled to one reservoir and weakly to others. 
        We emphasize that this model describes a general linear network of open oscillators, 
        and can appear in various physical systems \cite{Field_Locked_to_a_Fock_State,Cavity_optomechanics,Coherence_of_mechanical_oscillators,Establishing_Einstein_Poldosky_Rosen_Channels,Collective_Spins_in_Multipath_Interferometers,Effect_on_cavity_optomechanics}
        as well as model most Gaussian protocols in the CV quantum information 
        \cite{Quantum_information_with_continuous_variables,Gaussian_quantum_information}. 
        
        The state of $N$ oscillators can be completely described by the characteristic function defined as 
        $\chi(\bs{k},t)=\mathrm{Tr} \left[ \rho(t) \hat{D}(\bs{k}) \right]$. 
        Here $\hat{D}(\bs{k})= \exp \left[ \mathrm{i} (\bs{X}^T \bs{k}_x - \bs{P}^T \bs{k}_p) \right]$ 
        is the Weyl operator and 
        $\bs{k} = (\bs{k}_x , \bs{k}_p)$ 
        is a $2N$-component vector in the phase space \cite{Quantum_information_with_continuous_variables,Gaussian_quantum_information,Quantum_noise_matrix_for_multimode_systems}. 
        In this work we focus on Gaussian dynamics 
        and assume the initial state of the total system to be factorized 
        while each reservoir $H^{(\alpha)}$ has a thermal initial state with temperature $T^{(\alpha)}$. 
        Then the state evolution can be obtained by using the path-integral method 
        to the Feynman-Vernon influence functional \cite{feynman1963the19}, and satisfies that \cite{martinez2013dynamics24}
        \begin{equation}
            \label{evo_cf}
            \chi(\bs{k},t) \! = \! \chi(\bs{\Phi}(t) \bs{k},0)
            \exp \left[ \! - \frac{1}{2} \bs{k}^T \bs{\Sigma}(t) \bs{k} 
            + \mathrm{i} \bs{\Pi}^T(t) \bs{k} \right],
        \end{equation}
        where the transition matrix $\bs{\Phi}(t)$ and thermal covariance matrix $\bs{\Sigma}(t)$ 
        are $2N\times2N$ matrices, and $\bs{\Pi}(t)$ is a $2N$ vector. 
        These coefficients are determined by the Green function matrix $\bs{G}(t)$ of the system, 
        which satisfies 
        \begin{equation}
            \label{eq2}
            \ddot{\bs{G}}(t)+\bs{VG}(t)-2(\bs{\eta} \ast \bs{G})(t)=0
        \end{equation}
        with initial condition $\bs{G}(0)\!=\!0$ and $\dot{\bs{G}}(0)=\bs{I}$,
        where the symbol $\ast$ denotes the time convolution, i.e.,  
        $(A \ast B)(t)=\int_0^t \,\mathrm{d} \tau A(t-\tau)B(\tau)$. 
        Here $\bs{\eta}(t)= \int \,\mathrm{d}\omega \bs{I}(\omega) \sin \omega t$ 
        is the dissipation kernel depended on the environmental spectral density $\bs{I}(\omega)$, 
        where $\bs{I}(\omega)= \sum_\alpha \bs{I}^{(\alpha)}(\omega)$ 
        and $\bs{I}^{(\alpha)}(\omega)= \sum_k \bs{C}_k^{(\alpha)T} \bs{C}_k^{(\alpha)} \delta(\omega-\omega_k) /(2m_k \omega_k)$. 
        In this model, coefficients in Eq. (\ref{evo_cf}) are given by $\bs{\Pi}(t)=0$,  
        $\bs{\Phi}(t) \! = \! \left( \begin{array}{cc}
            \dot{\bs{G}}(t) & \bs{G}(t) \\
            \ddot{\bs{G}}(t) & \dot{\bs{G}}(t)
        \end{array} \right) $ and 
        $\bs{\Sigma}(t) \! \!  = \! \! \left( \begin{array}{cc}
            \bs{\sigma}^{(0,0)}(t) & \bs{\sigma}^{(0,1)}(t) \\
            \bs{\sigma}^{(1,0)}(t) & \bs{\sigma}^{(1,1)}(t)
        \end{array} \right) $, 
        where matrices $\bs{\sigma}^{(n,m)}$ are determined by 
        $\bs{\sigma}^{(n,m)}(t)=\int_0^t \,\mathrm{d}t_1 \int_0^t \,\mathrm{d}t_2
        \bs{G}^{(n)}(t_1) \bs{\nu}(t_1-t_2) \bs{G}^{(m)}(t_2)$ and the noise kernel 
        $\bs{\nu}(t) \! \!  = \! \!  \sum_\alpha \int \mathrm{d}\omega \bs{I}^{(\alpha)}(\omega)
        \coth (\omega/2k_BT_\alpha) \cos (\omega t)$ \cite{martinez2013dynamics24}.
        
        \emph{Existence conditions of localized modes.}---We 
        search for conditions that maintain the Green function $\bs{G}(t)$ 
        since dissipationless dynamics will appear 
        if the Green function $\bs{G}(t)$ shows a non-vanishing behavior, 
        otherwise the system would experience a full decay as seen from Eq. (\ref{evo_cf}). 
        Without loss of generality, we assume the coupling strength can be written as 
        $C_{ik}^{(\alpha)}=g C_{ik}^{'(\alpha)}$ 
        where $g$ is the global coupling strength and $C_{ik}^{'(\alpha)}$  
        describes the local microscopic details. 
        Meanwhile, we consider the environmental spectrum that contains a 
        gap between $0$ and $\omega_c$, i.e., the spectral density $\bs{I}(\omega)=\bs{0}$ 
        when $\omega$ is inside the gap (Fig.$\,$\ref{Fig1}(b)).
        Such a band-gapped spectrum can be achieved by 
        utilizing the environment engineering, e.g., photonic crystals \cite{Manipulation_of_dynamic_nuclear_spin_polarization,Control_of_spontaneous_emission,Quantum_electrodynamics_near_a_photonic_band_gap,Temperature_dependent_resonances_in_superconductor_photonic_crystal}.
        Results of other spectrums can also be obtained with small modifications to our theory.

        The matrix $\bs{G}(t)$ can be solved using the Laplace transform, 
        and we find 
        $\tilde{\bs{G}}(s)=[s^2\bs{I}+\bs{V}-2\tilde{\bs{\eta}}(s)]^{-1}$, 
        {\bf where the notation $\tilde{f}$ is used to denote the Laplace transformation of function $f$}. 
        We introduce the eigenvalue function 
        $\lambda_k(s,g) (k=1,2,\ldots,N)$, 
        which denotes the $k$th eigenvalue of 
        $\tilde{\bs{G}}^{-1}(s,g)=s^2\bs{I}+\bs{V}-2g^2 \tilde{\bs{\eta}}'(s)$. 
        Here $\tilde{\bs{\eta}}(s)= g^2 \tilde{\bs{\eta}}'(s)$. 
        The order of eigenvalues can be defined by using the eigendecomposition 
        $\tilde{\bs{G}}^{-1}(s,g)=\bs{U}(s,g) \bs{\Lambda}(s,g) \bs{U}^+(s,g)$, 
        where we fix the order of eigenvalues for specific $s_0, g_0$ 
        and requiring that both $\bs{\Lambda}(s,g)$ and $\bs{U}(s,g)$ 
        are continuous for $s$ and $g$. 
        This creates correspondences of eigenvalues between different $s$ and $g$. 
        Under this definition, the function 
        $\lambda_k(\mathrm{i}y,g)$ is monotonic decreasing for both 
        $y \in (0, \omega_c)$ and $g$ while $\lambda_k(x,g)$ 
        is monotonic increasing for $x \ge 0$ \cite{Supplimentary_material25}. 
        We note that the condition $\lambda_k(0,g)>0$ should be satisfied by physical systems 
        which can be properly described by the model in Fig.$\,$\ref{Fig1}(a). 
        If $\lambda_k(0,g)<0$ for specific $k$, 
        a real singular point will present and give rise to the divergent behavior in $\bs{G}(t)$. 
        {\bf
        Physically, as we will show later, a mode with imaginary frequency appears 
        when $\lambda_k(0,g)<0$. 
        Thus the total Hamiltonian has no ground state 
        and the system will tend to a state with infinite negative energy.
        }
        In this case the model is dynamical unstable 
        and should be corrected to avoid this unphysical feature 
        by including terms previously neglected, e.g., the nonlinear term. 
        Outside this unstable region, if we have $\lambda_k(\mathrm{i} \omega_c,g)<0$, 
        there must be one root $\omega_{sk}$ such that $\lambda_k(\mathrm{i} \omega_{sk},g)=0$. 
        This root is related to an imaginary singular point of $\tilde{\bs{G}}(s)$, 
        which produces a non-decay oscillation in $\bs{G}(t)$ with frequency $\omega_{sk}$, 
        and thus associates with dissipationless dynamics. 

        In fact, each imaginary singular point corresponds to a localized mode 
        with frequency inside the spectral gap. 
        This can be confirmed by diagonalizing the Hamiltonian into a set of normal modes, 
        where these modes are combinations of oscillator modes and bath modes, 
        and their position operators have the form 
        $\bs{X}'= \bs{u}^{(0)} \bs{X} +\sum_{\alpha} \bs{u}^{(\alpha)} \bs{Q}^{(\alpha)}$. 
        The $l$th column $\bs{u}_l^{(0)}$ of the matrix $\bs{u}^{(0)}$
        is determined by the secular equation 
        $(\bs{V}- \sum_{\alpha} \bs{C}^{(\alpha)T} \bs{V}_l^{(\alpha)-1} \bs{C}^{(\alpha)}) \bs{u}_l^{(0)T} = \omega^2_l \bs{u}_l^{(0)T}$, 
        where $(\bs{V}_l^{(\alpha)})_{jk} = (\omega_l^2-\omega_k^2) \delta_{jk}$ 
        and $\omega_l$ is the $l$th normal mode frequency \cite{Supplimentary_material25}. 
        This equation is equivalent to $\tilde{\bs{G}}^{-1} (\mathrm{i} \omega_l) \bs{u}_l^{(0)T} = \bs{0}$, 
        thus the singular point $\mathrm{i} \omega_{sk}$ creates a localized mode with frequency $\omega_{sk}$. 
        {\bf
        This result is quite different from 
        the bound-state behavior in finite dimensional systems, 
        where only bound states in low-excitation subspaces can be calculated 
        and dissipationless dynamics are only understood in the zero-temperature case.
        For Gaussian systems, bound states in the whole Hilbert space can be completely determined 
        and they formed localized modes which behave as free oscillators.
        During the dissipation process, 
        a significant amount of photons remain in localized modes 
        while others dissipate into the environment, 
        thus give dissipationless dynamics regardless of the reservoir temperature.} 
    
        Based on above discussions, 
        we can now formulate existence conditions of localized modes. 
        Firstly, if all reservoirs are neglected, the system behaves as 
        $N$ independent effective oscillators with effective frequencies $\omega_{0k}(k=1,\ldots,N)$ 
        given by square roots of eigenvalues of the coupling matrix $\bs{V}$. 
        If the frequency $\omega_{0k}$ is inside the gap 
        $(0, \omega_c)$, then the eigenvalue function $\lambda_k(\mathrm{i}y,g)$ 
        will possess a zero in $(0, \omega_c)$ since  
        $\lambda_k(\mathrm{i}\omega_{0k},g)<\lambda_k(\mathrm{i}\omega_{0k},0)=0$, 
        thus the $k$th localized mode appears for arbitrary coupling $g$. 
        Secondly, if the frequency $\omega_{0k}$ is
        outside the gap $(0, \omega_c)$, 
        there always exists a non-zero critical coupling $g_{ck}$ 
        which satisfies $\lambda_k(\mathrm{i}\omega_{c},g_{ck})=0$. 
        The corresponding localized mode presents 
        when the coupling strength $g$ exceeds the critical value $g_{ck}$ (Fig.$\,$\ref{Fig1}(c)). 
    
        \emph{Localized modes and dissipationless behavior.}---We 
        next focus on the system dynamics. 
        When the system-reservoir coupling is below 
        the smallest critical coupling, no localized mode can present 
        and the system experiences a complete decay. 
        In this case the matrix $\tilde{\bs{G}}(s)$ has no poles and we have
        \begin{equation}
            \bs{G}(t) \! = \! \bs{\mathcal{I}}(t) \! = \! -\frac{\mathrm{i}}{\mathrm{\pi}}
            \int^\infty_{\omega_c} \, \mathrm{d}y
            \sin yt \left( \bs{M}(y)-\bs{M}^+(y) \right),
        \end{equation}
        where $\bs{M}(y)$ is a $N \times N$ matrix. 
        Here we assume that the spectral density behaves as 
        $\bs{I}(\omega) \sim (\omega-\omega_c)^\alpha$, $\omega \rightarrow \omega_c$, 
        then the transient function $\bs{\mathcal{I}}(t)$ vanishes as 
        $O(t^{-1-\alpha})$ \cite{asymptoticapproximationsofintegrals26}. 
        Consequently, the transition matrix becomes zero and 
        the thermal covariance matrix evolves to 
        an asymptotic values determined by reservoirs. 
        The initial distribution finally disappears and 
        the system reaches the thermal equilibrium with the environment.
    
        Localized modes appear when the coupling strength 
        is above the corresponding critical coupling. 
        By including poles of $\tilde{\bs{G}}(s)$, 
        we obtain $\bs{G}(t)$ as
        \begin{equation}
            \bs{G}(t)=\bs{\Omega}(t)+\bs{\mathcal{I}}(t), \qquad 
            \bs{\Omega}(t)= \sum_k \bs{\Omega}_k \frac{\gamma_k}{\omega_{sk}}
            \sin \omega_{sk}t.
        \end{equation}
        Here $\bs{\Omega}(t)$ is the undamped oscillating term 
        associated with localized modes, 
        where $\bs{\Omega}_k$ is the $N \times N$ rank $1$ projection matrix 
        and the coefficient $\gamma_k$ is the residue 
        characterizes the amplitude of the $k$th localized mode. 
        The coefficient $\gamma_k$ is non-zero if and only if the coupling $g$ 
        is above the critical coupling $g_{ck}$. 
        Over long times, the transient function $\bs{\mathcal{I}}(t)$ vanishes 
        and only the undamped oscillation $\bs{\Omega}(t)$ remains. 
        Thus, the system dynamics in the long time limit can be described as
        \begin{gather}
            \bs{\Phi}(t) \sim \sum_k \frac{\gamma_k}{\omega_{sk}}
            \left( \begin{array}{cc}
                \omega_{sk} \cos \omega_{sk}t & \sin \omega_{sk}t \\
                -\omega_{sk}^2 \sin \omega_{sk}t & \omega_{sk} \cos \omega_{sk}t
            \end{array} \right)
            \otimes \bs{\Omega}_k, \nonumber \\
            \bs{\Sigma}(t) \sim \bs{\Sigma}_0 + \sum_{j,k} \bs{\Sigma}_{j,k}(t)
            + \sum_{j \neq k} \bs{\Sigma}_{j,-k}(t).
        \end{gather}
        where $\bs{\Sigma}_{j, \pm k}(t)=\bs{\Sigma}_{j, \pm k}^{(0)} \exp{[\mathrm{i}(\omega_{sj} \pm \omega_{sk})t]}+\mathrm{H.c.}$ and 
        $\bs{\Sigma}_0$ and $\bs{\Sigma}_{j, \pm k}^{(0)}$ are time independent coefficients. 
        Exact forms of coefficients 
        $\bs{M}(y)$, $\bs{\Omega}_k$, $\gamma_k$, $\bs{\Sigma}_0$, 
        and $\bs{\Sigma}_{j ,\pm k}^{(0)}$ are given in the Supplemental Material \cite{Supplimentary_material25}.
    
        Above results show that localize modes 
        give rise to dissipationless dynamics. 
        Each presented localized mode contributes a periodic rotation 
        in the phase space which is similar to the free evolution of the Wigner function. 
        Moreover, the thermal covariance matrix $\bs{\Sigma}(t)$ 
        no longer reaches an asymptotic value but oscillates with 
        various frequencies due to the interference between localized modes. 
        This implies that the system is far from the equilibrium 
        even in the long time limit.
    
        \emph{The weak-coupling limit.}---For 
        the weak coupling case, 
        where the oscillator-reservoir coupling is weak 
        compare to the effective frequency, such that 
        $g \ll \omega_{0k}$, the reservoir contribution 
        $-2\tilde{\bs{\eta}}(s)$ in the $\bs{G}^{-1}(s)$ can be viewed 
        as a small term so that the perturbation method can be applied. 
        The coupling matrix $\bs{V}$ can be diagonalized as 
        $\bs{V}_d=\bs{PVP}^+$, where $\bs{P}$ is a unitary matrix and 
        $(\bs{V}_d)_{jk}= \omega_{0k}^2 \delta_{jk}$. 
        Then we have 
        $\tilde{\bs{G}}_d^{-1}(\mathrm{i}y,g)= -y^2\bs{I}+\bs{V}_d-2g^2 \tilde{\bs{\eta}}_d'(\mathrm{i}y)$, 
        where $\tilde{\bs{G}}_d^{-1}= \bs{P} \tilde{\bs{G}}^{-1} \bs{P}^+$ and 
        $\tilde{\bs{\eta}}_d'= \bs{P} \tilde{\bs{\eta}}'^{-1} \bs{P}^+$. 
        In the non-degenerate case, such that 
        $g \ll |\omega_{sj} - \omega_{sk}|$ for $j \neq k$, 
        we can obtain first order corrections of eigenvalue functions as 
        $\lambda_k(\mathrm{i}y,g) = -y^2+\omega_{0k}^2-2g^2 \tilde{\eta}_{dkk}'(\mathrm{i}y)$. 
        This allows us to obtain localized mode frequencies and critical couplings as
        \begin{gather}
            \omega_{sk}^2 = \omega_{0k}^2 - g^2 \tilde{\eta}_{dkk}'(\mathrm{i} \omega_{0k}) 
            +O(g^4) \quad (\omega_{0k}<\omega_c), \nonumber \\
            g_{ck}^2= \frac{\omega_{0k}^2-\omega_c^2}{2\tilde{\eta}_{dkk}'(\mathrm{i} \omega_c)} \theta(\omega_{0k}-\omega_c) 
            + O((\omega_{0k}^2-\omega_c^2)^2).
        \end{gather}

        A clear physical picture can be drawn from this: 
        (i) the localized mode frequency $\omega_{sk}$ 
        is the effective frequency $\omega_{0k}$ 
        plus a negative shift of order $O(g)$, 
        and (ii) the critical coupling increases as $\omega_{0k}-\omega_c$ 
        when $\omega_{0k}$ is above and near the band edge. 
        Moreover, we can find that 
        $\gamma_k = 1+O(g^2)$ and 
        $\bs{U}(\mathrm{i} \omega_{sk})= \bs{I}+O(g^2)$ when $\omega_{0k}<\omega_c$. 
        Thus the decay process is prohibited and 
        the system is completely isolated from the environment 
        in the weak coupling limit if all 
        effective frequencies are inside the gap. 
        This is in contrast to the usual situations 
        where Markovian dynamics become {\bf dominating} 
        and the system experiences the exponential decay.
    
        \emph{Cavities in waveguides.}---To 
        illustrate our results, 
        we now present an example of a system with band-gapped environment. 
        Such a system can be experimentally realized in an array of coupled 
        cavities, which synthesized in optical waveguides \cite{Tight_Binding_Description_of_the_Coupled_Defect_Modes,verbin2013observation27,rechtsman2013photonic28,Single_photon_router,Controlling_the_transport_of_single_photons_by_tuning_the_frequency_of_either_one_or_two_cavities_in_an_array_of_coupled_cavities,Controllable_Scattering_of_a_Single_Photon_inside_a_One_Dimensional_Resonator_Waveguide,Composite_vortices_in_nonlinear_circular_waveguide_arrays,Two_component_Bose_Hubbard_model_in_an_array_of_cavity_polaritons,Few_cycle_spatiotemporal_optical_solitons_in_waveguide_arrays,Continuous_variable_entanglement_on_a_chip}.
        Here, each cavity $H_{Sn}= \omega_0 a_n^+ a_n$ 
        is linearly coupled to a waveguide 
        $H_{En}= \sum_k \omega_k b^+_{nk} b_{nk}$, 
        which consists of an array of linear defects. 
        For simplicity, we assume that both cavities and waveguides 
        are identical. The interaction Hamiltonian is 
        $H_{In}= \sum_k g_k (a_n^+ + a_n)(b^+_{nk} + b_{nk})$, 
        with $\omega_k= \omega_1 - 2\kappa \cos kx_0$ and 
        $g_k= \kappa_0 \sin kx_0$, where $\omega_1$ 
        is the frequency of linear defects, 
        $x_0$ and $\kappa$ are the spatial separation 
        and the hopping rate between adjacent defects respectively \cite{PhysRevA.96.023831.22,longhi2009spectral29}. 
        The coefficient $\kappa_0$ describes the coupling strength 
        between the cavity and its adjacent defect. 
        In the tight-binding approximation, we also consider the interaction 
        $\alpha \omega_0 (a_n^+ + a_n)(a_{n\pm1}^+ + a_{n\pm1})$ 
        between adjacent cavities $H_{Sn}$ and $H_{S(n\pm1)}$, 
        and the interaction 
        $\beta \sum_k g_k (a_n^+ + a_n) ( b^+_{(n\pm1)k} b_{(n\pm1)k} )$ 
        between the cavity $H_{Sn}$ and its adjacent waveguides $H_{E(n \pm 1)}$, 
        where $\alpha$ and $\beta$ are the relative coupling strength. 
        One can notice that the spectrum of 
        waveguides shows upper and lower frequency limits, 
        and there are two continuous gaps outside the energy band.
        \begin{figure}
            \centering
            \includegraphics[scale=0.110]{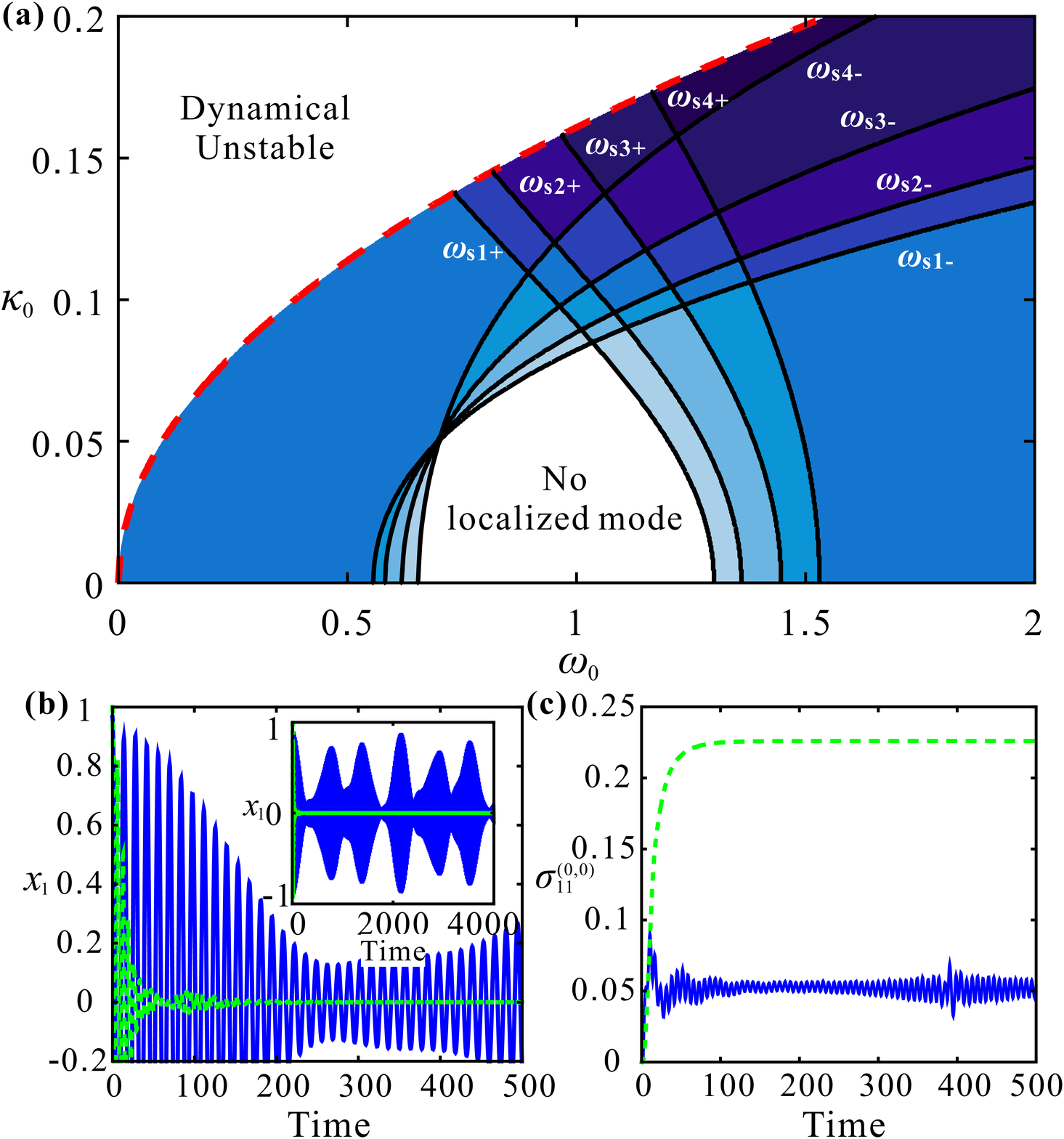}
            \caption{
                (a) Localized modes status of cavities in optical waveguides. 
                Each critical coupling line is labelled with the corresponding 
                mode frequency $\omega_{sk \pm}$, the sign $\pm$ indicates that 
                the localized mode is inside the band gap $(\omega_1+2\kappa,\infty)$ 
                or $(0,\omega_1-2\kappa)$. Localized modes can exist in the region 
                above their corresponding critical coupling lines. 
                (b) Time evolution of the expectation $\langle \hat{x}_1 \rangle$. 
                (c) Time evolution of the element $\sigma_{11}^{(0,0)}$ of the thermal covariance matrix. 
                In panels (b) and (c), blue solid lines (green dash lines) correspond to the circumstance 
                where localized modes present (absent), with $\kappa_0=0.05$ and $\omega_0=0.5$ ($\omega_0=1$). 
                The first cavity is initially in the coherent state $|\alpha \rangle$, 
                $\alpha =1$, while others are in vacuum states. 
                All reservoirs have vacuum initial states. 
                In all panels the system consists of four cavities, 
                while other parameters chosen are 
                $\omega_1=1$, $\kappa_0=0.05$, $\kappa=0.2$, 
                $\alpha=0.2$ and $\beta=0.2$.
            }
            \label{Fig2}
        \end{figure}
        Each gap can produce a group of localized modes, 
        and their existence is determined by matrices 
        $\tilde{\bs{G}}^{-1}(\mathrm{i} (\omega_1+2\kappa))$ 
        and $\tilde{\bs{G}}^{-1}(\mathrm{i} (\omega_1-2\kappa))$ respectively. 
        In Fig.$\,$\ref{Fig2}(a), we show the localized mode status 
        which agrees with our previous discussions clearly. 
        An unstable region exists in the upper left portion 
        and its extent is decided by $\tilde{\bs{G}}^{-1}(0)$. 
        Below this unstable region, the $k$th localized mode always exists 
        if the effective frequency $\omega_{0k}$ 
        is inside the environment band gap 
        $(0,\omega_0-2\kappa)$ or $(\omega_0+2\kappa,+\infty)$. 
        Moreover, there exists a critical coupling line for each localized mode, 
        which appears above this line.
    
        The system dynamics is shown in Figs.$\,$\ref{Fig2}(b) and \ref{Fig2}(c). 
        Clearly, when localized modes present, 
        the system experiences a partial decay and 
        becomes dissipationless after a long time. 
        In Fig.$\,$\ref{Fig2}(b), the time evolution of 
        $\langle \hat{x}_1 \rangle$ 
        can be described as the superposition of different periodic oscillations. 
        This indicates that in the long time limit, 
        the system behaves as the combination of several periodic rotations 
        in the phase space, 
        and each rotation corresponds to a localized mode. 
        Besides, Fig.$\,$\ref{Fig2}(c) shows that the system 
        fails to reach the equilibrium with the environment, 
        instead there exists periodic energy flows between them.

        Experimentally, localized modes can be produced by slightly detuning the cavity frequency $\omega_0$ 
        from the defect frequency $\omega_1$ by changing the geometrical parameters of cavities. 
        Since the waveguide has a very narrow band, 
        a small detuning is enough to insure the cavity frequency inside the band gap. 
        For example, in the coupled-cavity system in photonic crystal slabs, 
        the typical frequency of defects and hopping rate are 
        $\omega_1=0.305 \times 2\pi c/d$ and $\kappa= 1.5 \times 10^{-3} \times 2\pi c/d$, 
        where $d$ is the lattice period \cite{Engineering_the_quality_factors_of_coupled_cavity_modes_in_photonic_crystal_slabs}. 
        In order to probe the dynamics, we can directly measure photon currents flowing over waveguides. 
        These photon currents describe the tunneling of photons between cavities and waveguides, 
        and will oscillate persistently if localized modes present, 
        otherwise they will disappear rapidly 
        since the system will reach equilibrium with the environment in very short time. 
        Therefore, dissipationless dynamics can be confirm by the observation 
        of non-vanishing oscillating photon currents. 

        \emph{Conclusion.}---In summary, we have presented a general theory 
        of the dissipationless dynamics for open Gaussian systems, 
        and shown that dissipationless dynamics is a universal feature 
        for Gaussian systems with band-gapped environments. 
        This novel dynamics arises from localized modes, 
        which are formed as long as effective frequencies of oscillators 
        are inside the environmental spectrum gap or 
        the system-environment coupling exceeds the critical values. 
        Such a feature allows us to suppress environmental noises 
        by modifying the environment in order to induce localized modes. 
        Our theory can be applied to most Gaussian CV protocols, 
        and sheds light on the way to protect quantum resources in the CV quantum information. 
        It also provides a clue to understand the non-Markovianity 
        in more general many-body open quantum systems.

        This work was supported by NSFC under grants Nos. 11574022, 
        11434015, 61227902, 61835013, 11611530676, KZ201610005011,
        the National Key R\&D Program of China under grants Nos. 2016YFA0301500, 
        SPRPCAS under grants No. XDB01020300, XDB21030300.

        H. J. Z. and G. F. Z. contributed equally to this work.

        \bibliography{DissipationlessBehavior.bib}
    \end{document}